\documentclass[aps, prd, twocolumn, nofootinbib, 
floatfix,superscriptaddress]{revtex4-2}
\usepackage{subfigure}
\usepackage{epsfig}
\usepackage{amsmath}
\usepackage{amsfonts}
\usepackage{amssymb}
\usepackage{amssymb,amsmath,amsfonts}
\usepackage{xfrac}
\usepackage{color}
\pdfoutput=1
\usepackage[utf8]{inputenc}
\usepackage{graphicx}
\usepackage{dcolumn}
\usepackage{bm}
\usepackage{tikz}
\usepackage{amssymb}
\usepackage{float}
\usepackage{amsmath}
\usepackage{dcolumn}
\usepackage{cancel}

\usepackage{tcolorbox}
\usepackage{hyperref}
\hypersetup{colorlinks, citecolor=red, linkcolor=bluscuro, urlcolor=bluscuro}
\definecolor{rossos}{cmyk}{0,1,1,0.55}
\definecolor{bluscuro}{rgb}{0.15, 0.2, .85}
\definecolor{bluchiaro}{cmyk}{1,.3,0.,0.1}

\newcommand{\bea}{\begin{eqnarray}}
 
\newcommand{\eea}{\end{eqnarray}}
\newcommand{\bma}{\begin{pmatrix}}
\newcommand{\ema}{\end{pmatrix}}
\newcommand{\be}{\begin{equation}}
\newcommand{\ee}{\end{equation}}
\newcommand{\beno}{\begin{equation*}}
\newcommand{\eeno}{\end{equation*}}

\def\doi{http://doi.org}

\begin{document}

\title{A null test of the Hubble tension}

\author{Gerasimos Kouniatalis}
\email{gkouniatalis@noa.gr}
\affiliation{Physics Department, National Technical University of Athens,
15780 Zografou Campus,  Athens, Greece}
 \affiliation{Institute for Astronomy, Astrophysics, Space Applications and 
Remote Sensing, National Observatory of Athens, 15236 Penteli, Greece}

\author{Emmanuel N. Saridakis} \email{msaridak@noa.gr}
\affiliation{Institute for Astronomy, Astrophysics, Space Applications and 
Remote Sensing, National Observatory of Athens, 15236 Penteli, Greece}
\affiliation{Departamento de Matem\'{a}ticas, Universidad Cat\'{o}lica del 
Norte, Avda. Angamos 0610, Casilla 1280 Antofagasta, Chile}
 
\affiliation{CAS Key Laboratory for Researches in Galaxies and Cosmology, 
School 
of Astronomy and Space Science, University of Science and Technology of China, 
Hefei, Anhui 230026, China}


\begin{abstract}
The origin of the Hubble tension remains one of the central open problems in
modern cosmology, with competing explanations invoking either early-Universe
physics, late-time modifications of cosmic expansion, or unresolved
observational systematics. In this Letter we propose a new, purely geometric
null test of the late-time expansion history that is exactly independent of the
Hubble constant. By combining strong-lensing time-delay distances with
gravitational-wave standard-siren luminosity distances, we construct a
dimensionless ratio that depends only on the redshift dependence of the
expansion rate and can be both predicted from early-Universe data and measured
directly at late times, without relying on the cosmic distance ladder or the
sound horizon. We show that the comparison between the early- and late-time
determinations of this ratio provides a transparent consistency test of the
standard cosmological expansion. When combined with an independent
standard-siren measurement of $H_{0}$, this framework allows one to
unambiguously distinguish between early- and late-time origins of the Hubble
tension. With the forthcoming detection of lensed gravitational-wave standard
sirens, the proposed test provides a timely and robust framework for probing
this long-standing cosmological puzzle.
\end{abstract}

\maketitle

\section*{Introduction}

One of the most persistent challenges in contemporary cosmology is the 
so-called 
Hubble tension, namely the statistically significant discrepancy between the 
value of the present-day expansion rate of the Universe inferred from 
early-Universe observations and that obtained from late-Universe measurements. 
Within the standard flat $\Lambda$CDM paradigm, Cosmic Microwave Background 
observations by the Planck collaboration favor a value $H_0\simeq 67\,{\rm 
km\,s^{-1}\,Mpc^{-1}}$ \cite{Planck:2018vyg}, while local determinations based 
on the distance ladder yield a higher value $H_0\simeq 73\,{\rm 
km\,s^{-1}\,Mpc^{-1}}$ \cite{Riess:2019cxk}, with a discrepancy currently 
exceeding the $4$--$5\sigma$ level \cite{CosmoVerseNetwork:2025alb}.

A wide variety of approaches have been proposed in order to address this issue. 
These include extensions of the early-Universe sector, such as early dark 
energy 
or additional relativistic degrees of freedom 
\cite{Poulin:2018cxd,Simon:2022adh}, as well as late-time modifications of 
gravity or dark energy dynamics \cite{CANTATA:2021asi}. At the 
same time, considerable effort has been devoted to developing alternative 
observational strategies, including inverse distance ladders and time-delay 
strong lensing \cite{Cuesta:2014asa,Arendse:2019itb,Treu:2016ljm}, 
as well as gravitational-wave standard sirens 
\cite{Schutz:1986gp,LIGOScientific:2017adf}. Nevertheless, most existing 
analyses ultimately rely on either specific cosmological parametrizations or on 
measurements that are directly sensitive to the absolute expansion scale, 
making 
it difficult to disentangle whether the tension originates from the 
normalization of $H(z)$ or from its redshift dependence.

In this Letter we propose a complementary approach, based on a purely geometric 
and exactly $H_0$-independent null test of the late-time expansion history. By 
combining strong-lensing time-delay distances with gravitational-wave 
standard-siren luminosity distances, we construct a dimensionless observable 
that depends only on the shape of the Hubble function $H(z)$ and not on its 
overall normalization. This allows for a direct consistency test between 
early-Universe predictions, derived for instance within the Planck-preferred 
$\Lambda$CDM framework, and late-time observations, without invoking the cosmic 
distance ladder or the sound horizon. When this test is combined with an 
independent, low-redshift, determination of $H_0$ from standard sirens, it provides a clean 
diagnostic capable of distinguishing between early-time new physics, late-time 
departures from $\Lambda$CDM, and residual observational systematics. As such, 
the proposed framework offers a conceptually simple and robust tool that can be 
readily applied to forthcoming lensing and gravitational-wave data, and can 
play 
a valuable role in clarifying the physical origin of the Hubble tension.

\section*{Geometric construction}
\subsection*{Distances in FLRW geometry}

We consider a homogeneous and isotropic Universe described by a
Friedmann-Lemaître-Robertson-Walker (FLRW) geometry. In view of current 
observational
constraints, which are consistent with very small spatial curvature
\cite{Planck:2018vyg}, we restrict ourselves to the spatially flat case, for
which the line element reads
\begin{equation}
\mathrm{d}s^{2}=-c^{2}\mathrm{d}t^{2}
+a^{2}(t)\left(\mathrm{d}r^{2}+r^{2}\mathrm{d}\Omega^{2}\right),
\label{eq:flrw_flat}
\end{equation}
with $a(t)$ the scale factor and $\mathrm{d}\Omega^{2}$ the metric on the unit
two-sphere. Possible effects of spatial curvature will be briefly discussed
later.

The expansion rate is characterized by the Hubble parameter
$H(t)\equiv \dot a/a$, whose present value defines the Hubble constant
$H_{0}$. Introducing the redshift through $1+z=a^{-1}(t)$ and the dimensionless
expansion function $E(z)\equiv H(z)/H_{0}$, the comoving radial distance to a
source at redshift $z$ follows from null geodesic propagation as
\begin{equation}
\chi(z)=c\int_{0}^{z}\frac{\mathrm{d}z'}{H(z')}
=\frac{c}{H_{0}}\int_{0}^{z}\frac{\mathrm{d}z'}{E(z')}.
\label{eq:chi}
\end{equation}
For later convenience we define the dimensionless integral
\begin{equation}
I(z_{1},z_{2})\equiv\int_{z_{1}}^{z_{2}}\frac{\mathrm{d}z'}{E(z')},
\label{eq:Idef}
\end{equation}
such that $\chi(z)=\frac{c}{H_{0}}I(0,z)$.

The angular-diameter distance in a flat FLRW background is then given by
\begin{equation}
D_{A}(z)=\frac{\chi(z)}{1+z}
=\frac{c}{H_{0}}\frac{1}{1+z}I(0,z),
\label{eq:DA}
\end{equation}
while the luminosity distance is defined through the inverse-square law.
Assuming a metric theory of gravity and photon-number conservation, the
Etherington distance-duality relation holds \cite{Etherington:1933asu},
\begin{equation}
D_{L}(z)=(1+z)^{2}D_{A}(z),
\label{eq:duality}
\end{equation}
which leads to the compact expression
\begin{equation}
D_{L}(z)=\frac{c}{H_{0}}(1+z)\,I(0,z).
\label{eq:DL}
\end{equation}
Equation~\eqref{eq:DL} is purely geometric and valid for an arbitrary expansion
history $E(z)$, and thus all cosmological information is encoded in the 
dimensionless
integral $I(0,z)$, while the Hubble constant appears only as an overall
normalization. All quantities entering  \eqref{eq:DL} are defined 
geometrically, with no assumption on the matter content or dark-energy dynamics 
beyond the FLRW background and distance duality.

In the same geometric framework, strong-lensing time-delay observations define
the time-delay distance $D_{\Delta t}(z_{\rm d},z_{\rm s})$, which relates the
measured time delays between multiple images of a lensed source at redshift
$z_{\rm s}$ with lens redshift $z_{\rm d}$, to the angular-diameter distances between observer, lens, and
source. In a spatially flat FLRW background, this distance can be expressed as
\begin{equation}
D_{\Delta t}(z_{\rm d},z_{\rm s})
= \left(\frac{c}{H_0}\right)
  \frac{ I(0,z_{\rm d})\,I(0,z_{\rm s}) }
       { I(z_{\rm d},z_{\rm s}) } .
\label{eq:Ddt-final}
\end{equation}
As in the case of the luminosity distance, all dependence on the cosmological
model enters through the dimensionless expansion history $E(z)$, while the
Hubble constant appears only as an overall normalization.

\subsection*{An $H_0$-independent ratio}

We now combine the geometric luminosity distance of Eq.~\eqref{eq:DL} with the
strong-lensing time-delay distance of Eq.~\eqref{eq:Ddt-final}. Defining the
dimensionless ratio
\begin{equation}
R(z_{\rm d},z_{\rm s})
\equiv
\frac{D_{\Delta t}(z_{\rm d},z_{\rm s})}{D_{L}(z_{\rm s})},
\label{eq:Rdef}
\end{equation}
we obtain, after straightforward substitution,
\begin{equation}
R(z_{\rm d},z_{\rm s})
=
\frac{I(0,z_{\rm d})}{(1+z_{\rm s})\,I(z_{\rm d},z_{\rm s})}.
\label{eq:Rfinal}
\end{equation}

Equation~\eqref{eq:Rfinal} constitutes the central result of this work. Being
independent of the overall normalization of the expansion rate, $R$ provides a
purely geometric probe of the late-time cosmological background, sensitive only
to the redshift dependence of $H(z)$. In particular, any change in $H_{0}$ that
leaves the functional form of $E(z)$ invariant leaves $R$ unchanged. As a
result, this ratio enables a direct and model-agnostic consistency test between
early-Universe predictions for the expansion history and late-time geometric
measurements, without invoking the cosmic distance ladder or assuming a specific
dark-energy or modified-gravity parametrization, beyond the FLRW framework and
the validity of distance duality.

Finally, concerning spatial curvature, given the current observational 
constraints $|\Omega_k|\lesssim 3\times10^{-3}$,
any curvature-induced corrections to the ratio $R$ are expected to be at the
sub-percent level, and thus well below foreseeable observational uncertainties.

\section*{The null test and its interpretation}
 
\subsection*{Early- vs late-time consistency}

Within the standard cosmological framework, measurements of the Cosmic Microwave
Background (CMB) tightly constrain the parameters of flat $\Lambda$CDM 
scenario, and in
particular the fractional matter and dark-energy densities
\cite{Planck:2018vyg}. 
The null hypothesis underlying the present test is that the Universe is
described by a spatially flat FLRW geometry, distance duality holds, and the
late-time expansion history $E(z)$ follows the form implied by the CMB-fit
$\Lambda$CDM parameters.
Once these parameters are fixed, the expansion history
$E(z)$ is fully determined, and consequently the dimensionless integrals
$I(z_{1},z_{2})$ entering relation \eqref{eq:Rfinal} can be calculated 
unambiguously.
This allows one to obtain an early-Universe prediction for the ratio
$R(z_{\rm d},z_{\rm s})$, which we denote as $R_{\rm CMB}(z_{\rm d},z_{\rm s})$.
Importantly, since the Hubble constant cancels exactly in $R$, this prediction
is insensitive to the specific value of $H_{0}$ inferred from the CMB, and
depends only on the shape of the late-time expansion history implied by the
early-Universe data.

On the other hand, lensed gravitational-wave standard sirens provide a 
late-time and fully geometric determination of the same quantity. A strongly
lensed gravitational-wave event with an identified electromagnetic counterpart
yields the source redshift $z_{\rm s}$, the lens redshift $z_{\rm d}$, the
luminosity distance $D_{L}^{\rm GW}(z_{\rm s})$ from the waveform, and the
time-delay distance $D_{\Delta t}(z_{\rm d},z_{\rm s})$ from the measured delays
between multiple images and the reconstructed lens potential
\cite{Suyu:2012aa,Treu:2016ljm,Arendse:2019itb}. From these observables
one can directly form the empirical ratio
\begin{equation}
R_{\rm obs}(z_{\rm d},z_{\rm s})
\equiv
\frac{D_{\Delta t}^{\rm obs}(z_{\rm d},z_{\rm s})}
     {D_{L}^{\rm GW}(z_{\rm s})},
\label{eq:Robs}
\end{equation}
which, by construction, is independent of any distance ladder, sound-horizon
calibration, or prior on $H_{0}$.

Finally, to quantify the level of agreement between early- and late-time 
determinations,
it is convenient to introduce a dimensionless tension estimator,
\begin{equation}
T_R \equiv \frac{R_{\rm obs}(z_{\rm d},z_{\rm s}) -
R_{\rm CMB}(z_{\rm d},z_{\rm s})}{\sigma_R},
\end{equation}
where $\sigma_R$ denotes the total uncertainty on the observed ratio, 
propagated 
from the errors on $D_{\Delta t}^{\rm obs}$ and $D_{L}^{\rm GW}$.
A statistically significant deviation $|T_R|\gg 1$ directly signals a
breakdown of late-time consistency within the assumed cosmological framework.

\subsection*{Diagnosing the Hubble tension}

\begin{figure}[ht]
  \centering
  \includegraphics[width=0.46\textwidth]{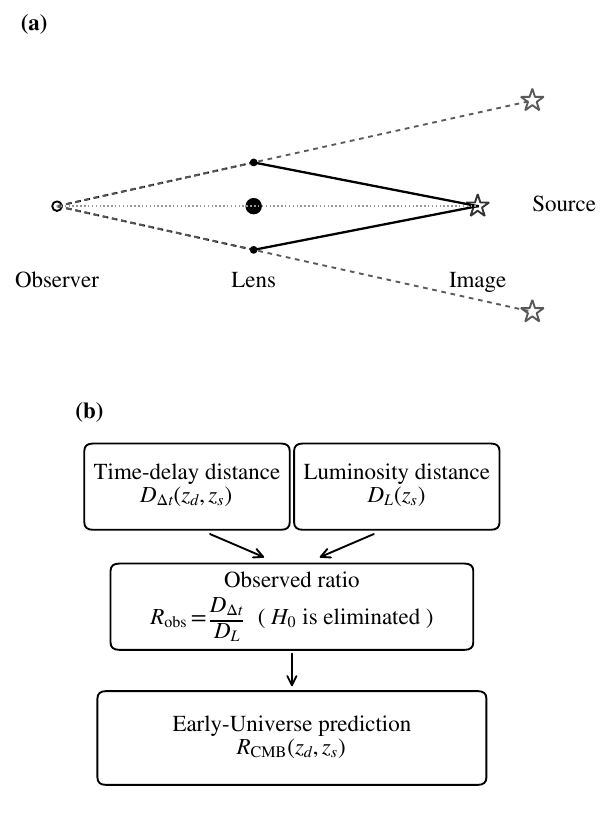}
  \caption{{\it{
Schematic overview of the proposed $H_0$-independent consistency test.
(a) Geometry of a strongly lensed gravitational-wave event: a lens at redshift
$z_{\rm d}$ produces multiple images of a source at $z_{\rm s}$.
The measured time delays determine the time-delay distance
$D_{\Delta t}(z_{\rm d},z_{\rm s})$, while the gravitational-wave amplitude 
yields
the luminosity distance $D_L(z_{\rm s})$.
(b) Conceptual structure of the null test: the observed ratio
$R_{\rm obs}=D_{\Delta t}/D_L$, in which the overall normalization of the
expansion rate is eliminated, is directly compared with the corresponding
early-Universe prediction $R_{\rm CMB}(z_{\rm d},z_{\rm s})$. 
}}}
\label{fig:schematic}
\end{figure}

The ratio constructed above allows for a direct and physically transparent
diagnosis of the origin of the Hubble tension. In particular, the null test proposed in this work consists in comparing the late-time,
observationally inferred $R_{\rm obs}$ with the early-Universe prediction
$R_{\rm CMB}$.  
In Fig. \ref{fig:schematic} we summarize the   geometric construction 
underlying our   test. 
In panel~(a) we illustrate the strong-lensing configuration that yields the 
time-delay distance 
$D_{\Delta t}(z_d,z_s)$, together with the luminosity distance $D_L(z_s)$ 
inferred from the amplitude 
of a (lensed) gravitational-wave standard siren. Additionally, panel~(b) shows 
how these two late-time observables 
are combined into the dimensionless ratio $R_{\rm obs}=D_{\Delta t}/D_L$, in 
which the dependence on 
the overall normalization of the expansion rate is eliminated, and how this 
empirical quantity can be 
directly confronted with the corresponding early-Universe prediction $R_{\rm 
CMB}(z_d,z_s)$.

Remarkably, all dependence on the Hubble constant cancels exactly, and the ratio
depends only on the redshifts of the lens and source, as well as on the shape of
the expansion history encoded in the dimensionless function $E(z)$. 
Importantly, $R$ is constructed entirely from directly measured quantities: the 
source and lens redshifts, the gravitational-wave luminosity distance, and the 
strong-lensing time-delay distance.

The diagnostic power of this test becomes manifest when it is combined with an
independent determination of the present-day expansion rate $H_0$ from gravitational-
wave standard sirens. Since, low-redshift standard sirens do not rely on the cosmic distance ladder
or on early-Universe physics, the comparison between $R_{\rm obs}$,
$R_{\rm CMB}$, and the standard-siren value of $H_{0}$   allows one to
disentangle inconsistencies in the shape of the expansion history from those in
its overall normalization.

If the ratio $R_{\rm obs}$ is found to be consistent with $R_{\rm CMB}$, while
the standard-siren determination of $H_{0}$ agrees with the CMB-inferred value,
then both the shape and normalization of the expansion history are mutually
consistent, and the Hubble tension may originate from residual systematics in
local distance-ladder measurements. Conversely, if $R_{\rm obs}$ remains
consistent with $R_{\rm CMB}$ but the standard-siren value of $H_{0}$ agrees 
with
local distance-ladder determinations instead, then the late-time expansion
geometry is unaltered and the tension points towards new physics operating
before recombination, capable of modifying the CMB-inferred Hubble scale while
leaving the late-time expansion history essentially unchanged. Finally, if
$R_{\rm obs}$ exhibits a statistically significant deviation from $R_{\rm CMB}$,
this directly signals a breakdown of the late-time $\Lambda$CDM expansion
history, favoring scenarios involving evolving dark energy, modified gravity,
or more general departures from the standard FLRW description.

In this way, the combination of the $H_{0}$-independent ratio $R$ with a direct
standard-siren measurement of $H_{0}$ provides a minimal yet powerful framework
for identifying whether the Hubble tension originates from early-Universe
physics, late-time modifications of cosmic expansion, or unresolved
observational systematics.

Finally, unlike inverse distance-ladder approaches based on BAO and supernovae, 
the 
present test relies exclusively on strong-lensing geometry and 
gravitational-wave standard sirens, without invoking the sound horizon or any 
external distance calibration.

\section*{Discussion and conclusions}

In this Letter we have introduced a new, purely geometric null test of the
late-time cosmological expansion, designed to directly probe the origin of the
Hubble tension. By combining strong-lensing time-delay distances with
gravitational-wave standard-siren luminosity distances, we constructed a
dimensionless ratio that is exactly independent of the Hubble constant. This
ratio depends only on the redshift dependence of the expansion rate and can be
predicted from early-Universe data, or measured directly at late times, without
invoking the cosmic distance ladder, the sound horizon, or any external
calibration. 

From an observational perspective, the proposed test is particularly timely in
view of the expected capabilities of third-generation gravitational-wave
detectors, such as the Einstein Telescope and Cosmic Explorer. Forecasts suggest
that these facilities will detect lensed gravitational-wave standard sirens at
rates of order $\mathcal{O}(1$--$10)$ per year, enabling the construction of
statistical samples rather than relying on individual events. The dominant
sources of uncertainty are expected to arise from strong-lens mass modeling,
which currently achieves percent-level precision in quasar time-delay studies,
as well as from gravitational-wave calibration and low-redshift peculiar
velocities affecting the luminosity-distance inference. While a single lensed
event will carry substantial uncertainties, a catalog of such systems will
allow these effects to be mitigated statistically, rendering the present
framework a viable and informative probe of late-time cosmological consistency.

A quantitative assessment of the statistical power of this test, for example
through Fisher forecasts or mock lensed-event catalogs for third-generation
gravitational-wave detectors, would require adopting detector- and
lens-model-specific assumptions and is therefore left for future dedicated
studies.

The central result is that the comparison between the observationally inferred
ratio and its early-Universe prediction provides a clean consistency test of
the late-time expansion geometry. When combined with an independent
standard-siren determination of $H_{0}$, this framework allows one to
unambiguously disentangle inconsistencies in the shape of the expansion history
from discrepancies in its overall normalization. In this way, the proposed test
offers a transparent diagnostic that can distinguish between early-Universe
solutions to the Hubble tension, late-time modifications of cosmic expansion,
and unresolved observational systematics.

An important feature of this approach is its minimal reliance on model
assumptions. Beyond the validity of the FLRW metric and distance duality, no
specific parametrization of dark energy, modified gravity, or early-Universe
physics is required. As a result, the test is broadly applicable and 
complementary
to existing probes, while remaining algebraically simple and conceptually
transparent.

We mention that possible theoretical systematics, such as modifications of 
gravitational-wave propagation, violations of distance duality, or small spatial 
curvature, would induce calculable corrections to the distances entering $R$. 
Nevertheless,  given current
constraints, these effects are expected to be subdominant but can be
straightforwardly incorporated if required.

With the rapid progress in time-delay cosmography and the advent of next-
generation gravitational-wave detectors, lensed standard sirens are expected to
become an observational reality in the near future. The framework developed
here provides a timely and well-defined way to fully exploit these observations,
offering a direct route towards probing one of the most persistent tensions
in modern cosmology.\\

{\bf{Acknowledgments}} --
  The authors acknowledge  the 
contribution of 
the LISA Cosmology Working Group (CosWG), as well as support from the COST 
Actions CA21136 -  Addressing observational tensions in cosmology with 
systematics and fundamental physics (CosmoVerse)  - CA23130, Bridging 
high and low energies in search of quantum gravity (BridgeQG)  and CA21106 -  
 COSMIC WISPers in the Dark Universe: Theory, astrophysics and 
experiments (CosmicWISPers).


\end{document}